# On the parameters of intermediate state structure in high pure type I superconductors at external magnetic field

Oleg P. Ledenyov, Vasiliy P. Fursa

*National Scientific Centre Kharkov Institute of Physics and Technology, Academicheskaya 1, Kharkov 61108, Ukraine.*

The geometric structure of an intermediate state in the high pure *Gallium* single crystal at the external magnetic field $H_{ext}$ at the temperature $T=0.4$ K is researched, using the longitudinal ultrasonic signal attenuation method at the ultrasonic signal frequency of *30 MHz*. It is experimentally shown that the edge inhomogeneities of the magnetic field distribution have an influence on the structure of an intermediate state in the cylindrical superconductor samples at the transverse orientation of external magnetic field $H_{ext}$. It is shown that the use of the superconducting plane screens of *NbZr* permits an approach to the equilibrium intermediate state structure of an infinite cylinder as confirmed by an experimental dependence of the normal metal layer thickness $a_N$ on the magnitude of external magnetic field $a_N(H_{ext})$ in the high pure *Ga* single crystal at an application of the longitudinal ultrasonic signal with the frequency of *30 MHz* at the temperature $T=0.4$ K. The experimentally obtained characteristic parameters of an intermediate state structure in the high pure *type I* superconductor at the external magnetic field $H_{ext}$ have a partial qualitative agreement with the *Landau* theory results.

## Introduction

The geometric structure of an intermediate state in the *type I* superconductor at the external magnetic field $H_{ext}$ is very sensitive to the magnetic field distribution and homogeneity in close proximity to the superconducting sample. In the case of the superconducting samples with complex geometrical forms, the topography of magnetic field strength lines, curving around the *type I* superconductor is complicated. The geometric structure of an intermediate state in the *type I* superconductor with the alternating superconductor - normal metal – superconductor (*S-N-S*) layers can also be very complicated and difficult to compute. There is a big number of possible variations of geometric structures of an intermediate state in the *type I* superconductors, which have the small free energies differences [1, p. 107]. The special experimental techniques have to be used to create an equilibrium ordered intermediate state structure in the *type I* superconductor in [2, 3]. The especially accurate detection and elimination of all the physical origins of non-homogeneities, resulting in an disappearance of the periodicity and homogeneity, are necessary in the case, when the research methods don't allow to make a direct visual observation of the *S-N-S* layers distribution in an intermediate state structure in the *type I* superconductor as in the case of the ultrasound attenuation research methods. The problem is that the short superconducting cylindrical samples with the demagnetization coefficient $\eta = 0.5$ were researched before, while the theoretical consideration was based on a model of the ideal infinitely long cylindrical superconducting sample with the demagnetization coefficient $\eta=0.5$, which was assumed to have a perfect structure of the alternating *S-N-S* layers in an intermediate state of the *type I* superconductor in [4]. The decrease of the magnitude of the demagnetization coefficient of a superconducting sample at the transverse external magnetic field $H_{ext}$ shows on the absence of cylindrical symmetry in the considered research problem, concerning the magnetic field distribution, which is evidently confirmed by the fact that the magnitude of demagnetization coefficient $n_z$ along the axe *z* is not equal to the nil, and the presence of concentration of magnetic field on the cylindrical superconducting sample's edges. It is possible to assume that the described physical problems on the magnetic field distribution are more related to the superconducting cylindrical sample's edges. The dependence of the demagnetization coefficient on the relative length of a cylindrical superconducting sample allows to limit the domains with the disordered intermediate state structure by the dimensions, corresponding to the order of magnitude of the radius of a cylinder near its edges. Thus, in the case of the cylindrical superconducting samples, which have the relation of the length *l* to the diameter *d* equal to the one, $l/d=1$, there is a symmetry in the magnetic field distribution, which is close to a spherical symmetry in the superconducting cylindrical samples with the demagnetization coefficient $n \simeq 1/3$ [5], hence there is no cylindrical symmetry in the considered research problem almost. The presence of regions with the different intermediate state structure with the various dimension and orientation differences in a superconducting cylindrical sample in comparison with the intermediate state structure in an ideal superconducting cylindrical sample, must have significant influence on the processes of an intermediate state structure formation during the possible change of magnitude of external magnetic field, resulting in a change of the periodicity of an intermediate state structure [6]. Moreover, the disappearance of an order in the *S-N-S* layers in the intermediate state structure in a superconducting sample must have an influence on both the nature of ultrasound attenuation and the choice of methodology of calculation of ultrasound attenuation in [7, 8], which was not considered before. The experimental researches [9, 12], which were conducted with the short superconducting samples ($\frac{l}{d} \leq 3$) with the application of the ultrasound attenuation method, found the dependence of the normal metal phase layer thickness $a_N$ and the period of structure *a* on the concentration of the normal metal phase $\eta$, which is close to the linear dependence at the $\eta > 0.5$, while there must be a sharp nonlinear increase of described dependence parameters in an agreement with the *Landau* theory results on an intermediate state structure in the high pure *type I* superconductors. It fully appears in the dependence of the normal metal layer thickness on the concentration of normal metal phase $a(\eta)$, which can be easy created during the researches with the application of the ultrasound attenuation method, using the data for the $a_N(\eta)$ and $a(\eta) = a_N(\eta)/\eta$. The difference between the experimental results and theoretical data doesn't permit to make a comparison with the calculations in [6] as it was already mentioned during the theoretical analysis [13], and the theoretical analysis with the use of the geometric resonance theory [14], considering the observed oscillations of ultrasound attenuation in an intermediate state in the high pure *type I* superconductor, observed in [10]. Assuming that the main problem is because of the described features of an intermediate state in the short cylindrical superconducting samples, we decided to make an attempt to solve the problem by improving the distribution of the magnetic field in close proximity to the edges of a superconducting cylindrical sample and by completing the experimental research toward the precise characterization of the main parameters of an intermediate state of the high pure *type I* superconductor.

## Precise characterization of parameters of intermediate state structure in high pure superconducting Gallium single crystal at external magnetic field

The ultrasound attenuation method has been used to research the geometric structure of an intermediate state in the high pure *type I* superconductors at the external magnetic field $H_{ext}$. This ultrasound attenuation method is similar to the ultrasound attenuation method in [15], where the research task on the "transformation" of a plane superconducting sample with the finite dimensions to a part of plane superconducting sample with the infinite dimensions (in the electrodynamics sense) was solved by the superconducting sample placing inside the superconducting cylindrical screen, which covered a superconducting sample tightly. In our case, during the consideration of problem with different symmetry, it is necessary to place a cylindrical superconducting sample with the finite dimensions between the two plane superconducting screens, which are in a tight contact with the cylindrical superconducting sample's edges, and perpendicular to the cylindrical superconducting sample's axe, and satisfy the following expression between the magnitudes of critical magnetic fields $H_{c1} > H_c$. The cylindrical symmetry of magnetic field distribution is recovered in a superconducting cylindrical sample with the modified geometry due to the use of the two plane superconducting screens. In this case, the superconducting cylindrical sample is a part of an infinite cylinder, going from the electrodynamics consideration.

The samples with the cylindrical geometric form with the diameter of *7 mm* and the length of *21 mm* were cut from the high pure *Ga* single crystal by the electro-erosion method. The axes of cylinders are oriented along the crystal axes *b* and *c*, and defined by the roentgenograph spectroscopy method with the precision *±1°*. The superconducting plane screens were made of the thin circular plates of alloy of *NbZr* with the thickness of *30-50 µm* and the diameter of *12-15 mm* with the magnitude of lower critical magnetic field $H_{c1} \approx 800\ Oe$ at the temperature of *4.2 K*, which were glued to the substrate made of glass-textolite. The special hole with the diameter of *4.5 mm* was made in the center of superconducting plane screens with the purpose to mount the ultrasonic transducers made of quartz plates with the *X* cut, designed to generate the main frequency of *10MHz*. The impulse method was used to conduct the ultrasonic measurements at the frequency of *30 MHz* in an analogy with experimental techniques in [10]. The cooling of the *Ga* samples and *NbZr* screens down to the *Helium* temperatures was conducted after the compensation of the *Earth* planet's magnetic field below the level $H_{residual} \lesssim 0.25\% H_{Earth}$, which was made by the *Helmholtz* inductive coils. The temperature of *0.4 K* was reached in the cryostat, using the cryo-adsorption pumping of the vapors above the fluid *Helium* three $^3He$. The magnitude of the critical magnetic field of the *Ga* single crystal was $H_c \approx 50\ Oe$, and it was much lower than the magnitude of the critical magnetic field of the superconducting plane screens of *NbZr*, $H_c \ll H_{c1}$, at the temperature of *0.4 K*. The analysis shows that, at the compensation of the magnetic field of the Planet, the magnitude of possible captured magnetic field in the superconducting plane screens of *NbZr* was so that the average (by the hole's cross-section) magnetic field was well below $\sim 10^{-4}\ H_c$. This magnitude of average magnetic field was in the three orders of magnitude smaller than the magnitude of external magnetic field along the main axe of the superconducting cylindrical sample near to its edges during the experiment without the superconducting plane screens.

The results of experiments are presented in Figs. 1, 2, 3. The decrease of value of the demagnetization coefficient down to $n \cong 0.44 - 0.42$ is characteristic in the experiments without the superconducting plate screens similar to the reported cases [5, 9-12]. It results in the bias of the superconductor – intermediate state transition point to the region of big magnetic fields in Fig. 1. In the experiments with the superconducting plane screens, the value of the demagnetization coefficient increased up to $n \cong 0.5$ in Fig. 1 (*B*). Let us evaluate the difference between the states in the first case and the second case from the thermodynamics point of view. The work, which is necessary to change the magnitude of the demagnetization coefficient, can be written as in the following expression

$$W = - \int_{n_{sample}}^{n_{ideal}} \frac{\partial M_i}{\partial H_e} \partial n .$$

The magnitude of this work characterizes an order of difference between the researched superconducting cylindrical sample with the finite dimensions and the ideal cylindrical superconducting sample with the infinite dimensions. The magnitude of work is equal to $W \approx 10 Erg$ in the case of a superconducting cylindrical sample with the finite dimensions without the superconducting plane screens at the external magnetic field $H_{ext} \approx 0.6\ H_c$ at the value of the total energy of superconducting state on the unit of superconductor volume $E \cong H_c^2/8\pi \cong 100\ Erg/cm^3$. Let us note that the surface energy of the phases separation boundaries near to the place with the maximal number of *S-N-S* layers in an intermediate state of the high pure *type I* superconductors, which can be evaluated going from the results [16, 17], is below *1 Erg* per a sample even at the concentration of the normal metal phase $\eta \approx 0.5$. Let us remind that the minimization of the value of surface energy and the minimization of the exit energy of the *S-N-S* layers define the structure of an intermediate state of the high pure *type I* superconductors Thus, the value of surface energy, is on the order of magnitude smaller than the value of magnetic energy, which is connected with the edges effects in an intermediate state of the high pure *type I* superconductors.

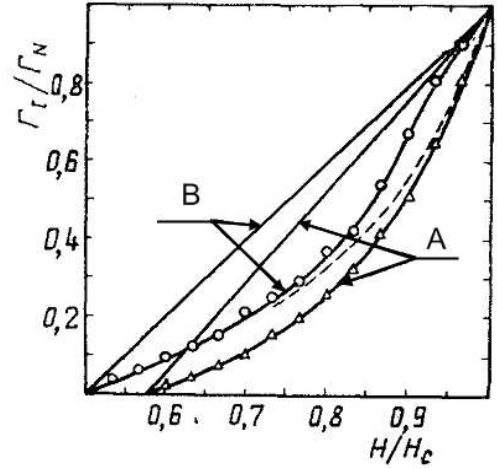

*Fig. 1. Dependence of relative ultrasound attenuation $\Gamma_l / \Gamma_N$ as function of external magnetic field $H / H_C$ in high pure Ga single crystal at temperature T=0.4 K, when longitudinal ultrasonic signal frequency is 30MHz, **k** // **c**, **H** // **a**: 1) no screens (A); 2) screens (B); 3) possible shape of curve of dependence $\Gamma_l / \Gamma_N$ on $H / H_C$ in case (B), if $(a_N)_B \cong (a_N)_A$ (dotted line).*

In the Figs. 2 and 3, the dimensions of layers with the normal metal phase in the cases of the two crystal orientations in an intermediate state of the high pure *Ga* single crystal are shown. These dimensions of layers with the normal metal phase were found, using the dependence of the ultrasound attenuation with the application of the theory [4]. In the region with the concentration of the normal metal phase $\eta > 0.7$, there is a considerable difference between the physical behaviors of the normal metal layers thicknesses in the following two cases:
1) the high pure *Ga* single crystal without the superconducting plane screens, made of *NbZr*, at the external magnetic field $H_{ext}$ (the lower curves (*A*) in Figs. 2 and 3);
2) the high pure *Ga* single crystal with the superconducting plane screens, made of *NbZr*, at the external magnetic field $H_{ext}$ (the upper curves (*B*) in Figs. 2 and 3).

In the second case, the sharp increase of the magnitude of the normal metal layer thickness $a_N$ is observed in every of the two cases of the crystal orientations in an intermediate state of the high pure *Ga* single crystal. The obtained research results show that the use of the diamagnetic screens during the experimental research with the application of the ultrasound attenuation method permits an approach to the equilibrium structure of an infinite cylinder in the short superconducting samples similar to the above described case of plane plate [15]. This fact is confirmed by the nature of the dependence of the normal metal layer thickness on the concentration of normal metal phase $a_N(\eta)$, which is in a partial agreement with the theory [6] in distinction from the experimental results, obtained with the use of the ultrasound attenuation method, and conducted without the improvement of an intermediate state structure parameters in the high pure *type I* superconductors. The small absolute value of the normal metal layer thickness $|a_N|$ in the region with the concentration of the normal metal phase $\eta < 0.7$ attracts certain attention. In our opinion, this research result can't be interpreted as an experimental result, which is in an agreement with the theory [6], considering both the models with the changed intermediate state structure or the renormalization of the function $\Phi\left(\dfrac{ka_N}{\pi}\right)$ in [4].

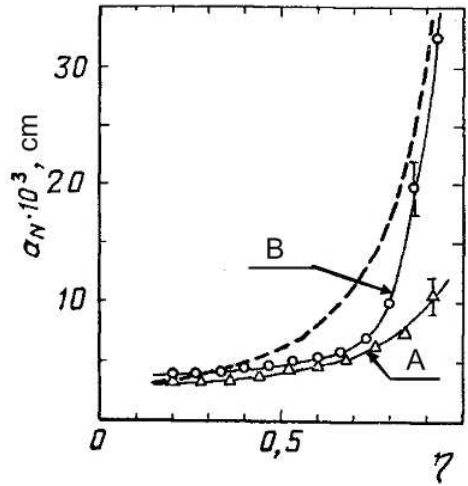

*Fig. 2. Dependence of normal metal phase layer thickness $a_N$ as function of normal metal phase concentration $\eta$ in high pure Ga single crystal at temperature T=0.4K, when longitudinal ultrasonic signal frequency is 30MHz, **k** // **c**, **H** // **a**: 1) no screens (A); 2) screens (B); 3) Landau theory ($a_{N\,th} = a_{N\,exp}$ at $\eta = 0,3$) (dotted line).*

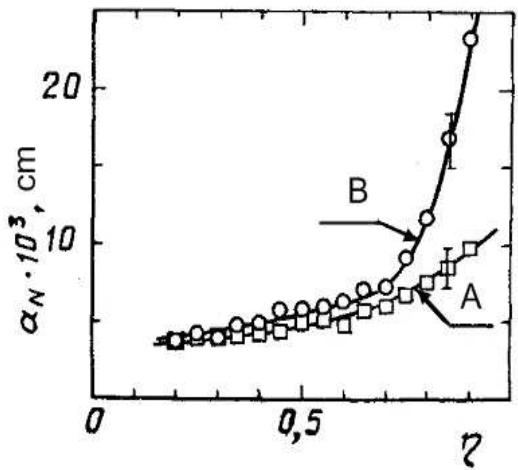

*Fig. 3. Dependence of normal metal phase layer thickness $a_N$ as function of as function of normal metal phase concentration $\eta$ in high pure Ga single crystal at temperature T=0.4K, when longitudinal ultrasonic signal frequency is 30MHz, $\angle$**H**, **c** = 22°, **k** // **b**: 1) no screens (A); 2) screens (B).*

In our opinion, the additional research is necessary to understand the nature of physical processes.

## Conclusion

The geometric structure of an intermediate state in the high pure *type I* superconductors at the external magnetic field $H_{ext}$ is researched by the ultrasound attenuation method. The advanced measurement set up to accurately characterize the geometric structure of an intermediate state in the high pure *Ga* single crystal at the frequency of *30 MHz* at the temperature of *0.4 K* is created. It is experimentally shown that the edge inhomogeneities of the magnetic field distribution have an influence on the geometric structure of an intermediate state in the high pure *type I* superconductor with the cylindrical shape. The proposed application of the superconducting plane screens of *NbZr* permits an approach to the equilibrium structure of an infinite cylinder as confirmed by the nature of an experimental dependence of the normal metal layer thickness $a_N$ upon the external magnetic field $a_N(H_{ext})$ in the high pure *Ga* single crystal at the frequency of *30 MHz* at the temperature T=0.4K. The experimentally obtained characteristic parameters of an intermediate state structure in the high pure *type I* superconductor have a partial qualitative agreement with the *Landau* theory results, when $\eta \to 1$ only.

Authors sincere thank Boris G. Lazarev for the useful thoughtful discussions on the innovative experimental research results; Alexander F. Andreev for the numerous stimulating encouraging discussions on the theoretical and experimental research results; Lyubov S. Lazareva for the superconducting plane screens design and some valuable research advises on the configuration of experimental set up, precise calibration of measurements devices, and accurate characterization of experimental results.

This research paper was published in the *Journal of Low Temperature Physics* (*FNT*) in 1985 [18].

*E-mail:   ledenyov@kipt.kharkov.ua